\documentclass[aps,12pt,preprint,,tightenlines,showkeys]{revtex4}
\usepackage[dvips]{color,graphicx}

\def\be{\begin{eqnarray}}
\def\ee{\end{eqnarray}}
\def\bea{\begin{eqnarray}}
\def\eea{\end{eqnarray}}
\def\beas{\begin{eqnarray*}}
\def\eeas{\end{eqnarray*}}
\newcommand{\eq}[1]{Eq.~(\ref{#1})}

\def\bfb{{\bf b}}

\def\bfR{{\bf R}}

\def\bfq {{\bf q}}
\def\bfv {{\bf v}}

\def\bfk{{\bf k}}

\def\bfp{{\bf p}}
\begin{document}


\title{CHARGE DENSITIES OF THE NEUTRON AND PROTON}
\author{
Gerald   A. Miller}

\affiliation{Department of Physics,
University of Washington\\
Seattle, Washington 98195-1560
}

\begin{abstract}
A model-independent analysis of the  infinite-momentum-frame 
charge density of partons in the transverse plane is presented 
for the nucleon. We find that the neutron parton charge density 
is negative at the center, so that the square of the transverse 
charge radius is positive, in contrast with many expectations. 
Additionally, the proton's central $u$ quark charge density is 
larger than that of the $d$ quark by about 70\%. The proton (neutron) 
charge density has a long range positively (negatively) charged 
component.

\end{abstract}
\maketitle

A truly impressive level of experimental technique, effort and ingenuity has been
brought to   measuring  the electromagnetic form factors
of the proton and neutron (nucleon) \cite{reviews}.
 These quantities are  probability amplitudes
that the  nucleon can absorb a given amount of momentum and  remain in the ground
state, and as such should be important sources of information about the nucleon
charge and magnetization densities.

The  text-book  interpretation of these form factors is  that their
 Fourier transforms are measurements of the charge and magnetization densities.
But  the initial and final nucleons have different momentum, and therefore different wave functions.
This is because the  
relativistic boost operator that  transforms a nucleon at  rest  into  a moving one changes
the wave function in a manner that depends on the momentum of the nucleon. 
The presence of different wave functions of  the initial and final nucleons invalidates  a
  probability or density  interpretation. 

A proper determination of a 
 charge density requires that the quantity be related
to the square of a wave function or of a field operator.
The technical solution to the problem of determining the relevant density operator
 has been known
for  a long  time \cite{soper1}, and  has been elegantly explained recently\cite{mbimpact,diehl2}
in terms of generalized parton distributions GPDs. 
The charge
density
$\rho(\bfb)$ \cite{notation}            of partons in the transverse plane is a  two-dimensional Fourier 
transform of the $F_1$ form factor. Here we present
 the first 
 phenomenological analysis of existing data to determine
$\rho(\bfb)$ for the neutron and proton. The results for the neutron
contradict the long-standing  notion, derived from
both gluon-exchange and meson-cloud models \cite{isgurk,cbm}, that the non-vanishing charge density 
at the center of the neutron 
is positive. 

 We begin by  by presenting    definitions of the  form
factors. Let $J^\mu(x)$ be the electromagnetic current operator, in units of the proton 
charge. Then  the nucleon form factors are given by 
\bea
\langle p',\lambda'| J^\mu(0)| p,\lambda\rangle =\bar{u}(p',\lambda')\left(\gamma^\mu F_1(Q^2)+i{\sigma^{\mu\alpha}\over 2M}q_\alpha F_2(Q^2)
\right) u(p,\lambda),\eea
where the momentum transfer $q_\alpha=p'_\alpha-p_\alpha$ is taken as space-like, so that 
$Q^2\equiv -q^2>0.$ The nucleon polarization states are chosen to be those of 
definite light-cone helicities $\lambda,\lambda'$ \cite{soper}.  
The charge (Dirac) form factor is $F_1$, normalized such that $F_1(0)$ is the
nucleon charge, and the magnetic (Pauli) form factor is $F_2$, normalized such that $F_2(0)$ is the
anomalous magnetic moment. the Sachs form factors\cite{sachs}
\bea
G_E(Q^2)\equiv F_1(Q^2)-{Q^2\over 4M^2}F_2(Q^2),\; G_M(Q^2)\equiv F_1(Q^2)+F_2(Q^2),\label{sachsff}\eea
were introduced so as  to provide an expression for the electron-nucleon cross section
(in the one photon exchange approximation) that depends on the quantities $G_E^2$ and $G_M^2$ but not
the product $G_E\;G_M$.
In  the Breit frame, in which 
$\bfp=-\bfp'$,  $G_E$ is the nucleon helicity flip matrix element of $J^0$. Furthermore, 
 the scattering of neutrons from the electron cloud of atoms  measures
the derivative $-d G_E(Q^2)/dQ^2$ at $Q^2=0$, 
 widely interpreted as six times
the mean-square charge radius of the neutron.  However,
any probability or density interpretation of $G_E$ 
is spoiled by a non-zero  value of $Q^2$, no matter how small \cite{sachsexplain}.
Any attempt to analytically incorporate relativistic corrections in a 
$p^2/m_q^2$ type of expansion  would be doomed, by the presence of the 
quark mass, $m_q$, to  be model-dependent.

The newly appreciated and  widely studied generalized 
parton distributions GPDs \cite{vdh}
are of high current interest because
 they 
can be related to the total angular
momentum carried by quarks in the nucleon and can be determined using
deeply virtual Compton scattering experiments \cite{xdj}.  These distributions are specific  matrix elements of quark field operators,
between nucleon states, which in contrast to the usual quark distribution functions,
 do not have the same momenta. We consider the specific case in which
the longitudinal momentum transfer $\xi$ is zero, and $\lambda'=\lambda$. Then,  
in the light-cone gauge, $A^+=0$, the matrix element defining the GPD, $H_q$ for a quark of flavor $q$ \cite{scale}
is
\bea&& 
{ H}_{q}(x,t)
=\int\!\! \frac{dx^-}{4\pi}\langle p^+,\bfp',\lambda|
\bar{q}(-\frac{x^-}{2},{\bf 0})
\gamma^+ q(\frac{x^-}{2},{\bf 0})
|p^+,\bfp,\lambda\rangle e^{ixp^+x^-}.
\label{eq:pd}
\eea
We use the abbreviation $H_q(x,\xi=0,t)\equiv H_q(x,t)$
and $ -t=-(p'-p)^2=(\bfp'-\bfp)^2=-q^2=Q^2.$
The simple form of $t$ results from its invariance under transverse boosts \cite{soper}:
Lorentz transformations, defined by a transverse vector $\bfv$
 that transform a four-vector $k$ according to 
$k^+\rightarrow k^+,\;\bfk\rightarrow\bfk-k^+\bfv$ and $k^-$ such that $k^2$ is unchanged.
 These quantities are part of a kinematic subgroup of the 
Poincar\'{e} group that obey the same commutation relations as those among the generators of
the Galilean transformations for non-relativistic quantum mechanics in the transverse plane.
The presence of the operator $\gamma^+$ insures that independent field operators appear 
in the matrix element.

GPDs allow for a unified description of a number of
hadronic properties \cite{xdj}. The most relevant for us are
that if $t=0$ 
 they reduce to conventional PDFs
$
H_q(x,0)= q(x)$, and that the 
  integration of $H_q$ over $x$ yields the nucleon 
electromagnetic  form factor:
\be
F_1(t)=\sum_q e_q \int dx H_q(x,t).  
\label{eq:form}
\ee

The spatial structure of a  nucleon can be examined if
one  uses \cite{soper1,mbimpact,diehl2} nucleonic 
states that are transversely localized.  The state with transverse center of mass
$\bfR$ set to 0 is formed by taking a  linear superposition of
states of transverse momentum.
In particular,
\be
\left|p^+,{\bf R}= {\bf 0},
\lambda\right\rangle
\equiv {\cal N}\int \frac{d^2{\bf p}}{(2\pi)^2} 
\left|p^+,{\bf p}, \lambda \right\rangle.
\label{eq:loc}
\ee
where $\left|p^+,{\bf p}, \lambda \right\rangle$
are light-cone helicity eigenstates
\cite{soper} and
${\cal N}$ is a normalization factor satisfying
$\left|{\cal N}\right|^2\int \frac{d^2{\bf p}_\perp}{(2\pi)^2}=1$.
References~\cite{mb1,diehl} use  
wave packet treatments that  avoid states 
normalized to $\delta$ functions, but  this  leads to the
same results as using \eq{eq:loc}. Note however,  the relevant range of integration
in \eq{eq:loc} must be restricted to $|\bfp|\ll p^+$ to maintain the interpretation
of a nucleon moving with well-defined longitudinal momentum\cite{mb1}. Thus we use
a frame with very large $p^+$. It is in just such a frame that 
the interpretation of a nucleon as a  set of a large number of partons is valid.  

Using  \eq{eq:loc} sets 
the  transverse 
center of momentum of 
a state of  total very large 
momentum $p^+$  to zero, so that
transverse distance $\bfb$ relative to $\bfR$.
can be  defined. 
Thus we may  generalize the quark-field operator appearing in  \eq{eq:pd}
 by making a translation:
\be
\hat{O}_q(x,{\bf b}) \equiv
\int \frac{dx^-}{4\pi}{q}_+^\dagger
\left(-\frac{x^-}{2},{\bf b} \right) 
q_+\left(\frac{x^-}{2},{\bf b}\right) 
e^{ixp^+x^-}.
\label{eq:bperp}
\ee The  
impact parameter dependent PDF is defined \cite{mb1} as the matrix element
of this operator in the state of \eq{eq:loc}:
\be
q(x,{\bf b}) \equiv 
\left\langle p^+,{\bf R}= {\bf 0},
\lambda\right|
\hat{O}_q(x,{\bf b})
\left|p^+,{\bf R}= {\bf 0},
\lambda\right\rangle. 
\label{eq:def1}
\ee

The use of \eq{eq:loc} in \eq{eq:def1} allows one to show that 
$q(x,{\bf b})$ is the two-dimensional Fourier transform of the GPD $H_q$:
\bea q(x,{\bf b})=\int {d^2q\over (2\pi)^2}e^{i\;\bfq\cdot\bfb}H_q(x,t=-\bfq^2),\label{ft1}
\eea with $H_q$ appearing because the initial and final helicities are each $\lambda$.
A complete determination of $H_q(x,t)$ (with $t\le0$) would  determine
$q(x,{\bf b})$.

One finds a probability interpretation \cite{soper1} by integrating $q(x,{\bf b})$
over all values of $x$. This  sets the value of $x^-$ to 0, so that 
\be
\int dx\;q(x,{\bf b}) \equiv 
\left\langle p^+,{\bf R}= {\bf 0},
\lambda\right|q_+^\dagger(0,\bfb)
q_+(0,\bfb)
\left|p^+,{\bf R}= {\bf 0},
\lambda\right\rangle.
\label{eq:def2}
\ee
If one multiplies the above relation   by  the quark charge $e_q$ (in units of $e$),
sums over quark flavors,  uses \eq{eq:loc} with
$\hat{O}_q(x,{\bf b})=e^{-i\hat{\bfp}\cdot\bfb}\hat{O}_q(x,{\bf 0}) e^{i\hat{\bfp}\cdot\bfb}$ 
 along with \eq{eq:form}, the resulting infinite-momentum-frame  IMF parton
charge density in transverse space
is 
\bea
\rho(b)\equiv \sum_q e_q\int dx\;q(x,{\bf b})=\int {d^2q\over (2\pi)^2} F_1(Q^2=\bfq^2)e^{i\;\bfq\cdot\bfb}.
\label{rhob}\eea

This transverse charge density $\rho(b)$, based on 
 $F_1$  \cite{soper1}, seems to  
contrast with the lore
relating  the charge density to 
$G_E$. There is no conflict, a feature
 understood by considering the relation between the Breit frame and 
the IMF. In the  Breit frame, the helicity
flip matrix element of $J^0$ is $G_E$. One uses a Lorentz transformation to 
consider the form factor  in the IMF. 
Then  the operator helicity-flip matrix element of $J^0$  becomes
the helicity non-flip matrix element of 
$J^+$, or  $F_1$. The transformation to the IMF
gains model independence  while  replacing $G_E$  by $F_1$.

We
exploit \eq{rhob} by  using measured form factors
to determine $\rho(b)$. Recent parameterizations \cite{Bradford:2006yz,Kelly:2004hm,
Arrington:2003qk} of $G_E$ and $G_M$ are very useful so we use \eq{sachsff} to
obtain $F_1$ in terms of $G_E,G_M$. Then
$\rho(b)$ can be expressed as a simple  integral
of known functions:
\bea \rho(b)= \int_0^\infty\; {dQ\;Q\;\over  2 \pi}J_0(Q b) {G_E(Q^2)+\tau G_M(Q^2)\over 1+\tau},\label{use}\eea
with $\tau={Q^2\over 4M^2}$ and $J_0$ a cylindrical Bessel function.

\begin{figure}
\unitlength.9cm
\begin{picture}(14,8.2)(1.5,2.4)
\includegraphics{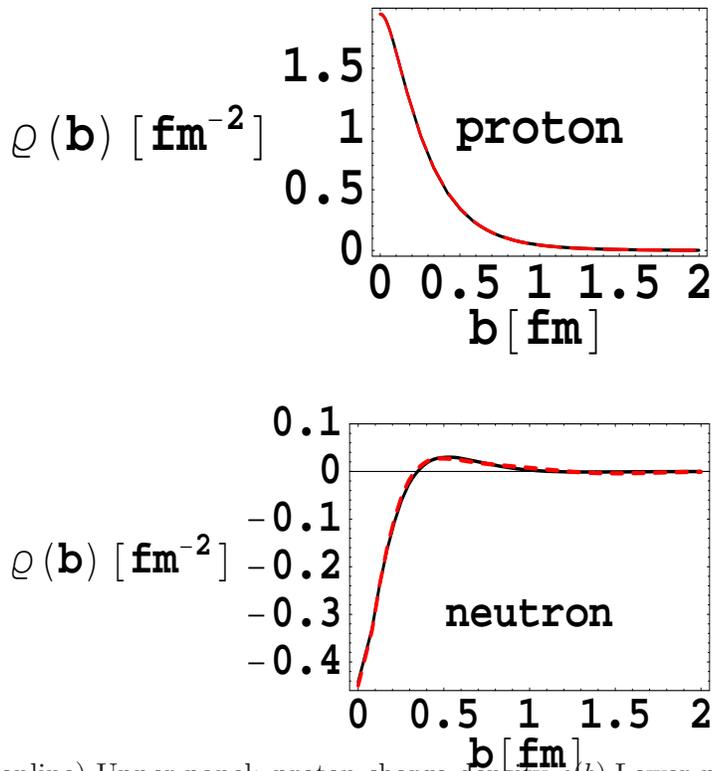}
\end{picture}\label{proton}
\caption{(Color online) Upper panel: proton charge density $\rho(b)$.Lower panel: neutron charge density. The solid curves use the  parameterization of
\cite{Kelly:2004hm}, and the dashed (red) curve uses \cite{Bradford:2006yz}. }
\end{figure}
A straightforward application of \eq{use} to the proton 
using the parameterizations \cite{Bradford:2006yz,Kelly:2004hm}
yields the results shown in the upper panel of  Fig.~1. The curves obtained using the two different parameterizations overlap. Furthermore, there is
negligible sensitivity to form factors at very high values of $Q^2$ that are currently unmeasured. The density is peaked 
at low  values of $b$, but has a long positive tail, suggestive of a long-ranged, positively charged 
  pion cloud.

The neutron  results 
are shown in the lower panel of 
 Fig.~1. The curves obtained using the two different parameterizations seem to overlap, 
but see below. The surprising result
is that the central 
neutron charge density is negative. If the neutron is sometimes a proton surrounded by a negatively
charged pionic cloud, one would expect to obtain a positive central density\cite{cbm}. 
Another  mechanism involving correlations in the
nucleonic wave function induced by  one gluon
exchange would also  lead to a positive central density
because  the interaction between two identical $d$ quarks\cite{isgurk} is 
 repulsive.  The values of the integral of \eq{use} are somewhat sensitive to the
regime $2<\tau<4$ for which $G_E$ is  as yet   unmeasured. About 30\% of the value of 
$\rho(0)$ arises from this region.

\begin{figure}\label{neutron}
\unitlength1.cm
\begin{picture}(14,8.2)(2.5,1.9)
\includegraphics{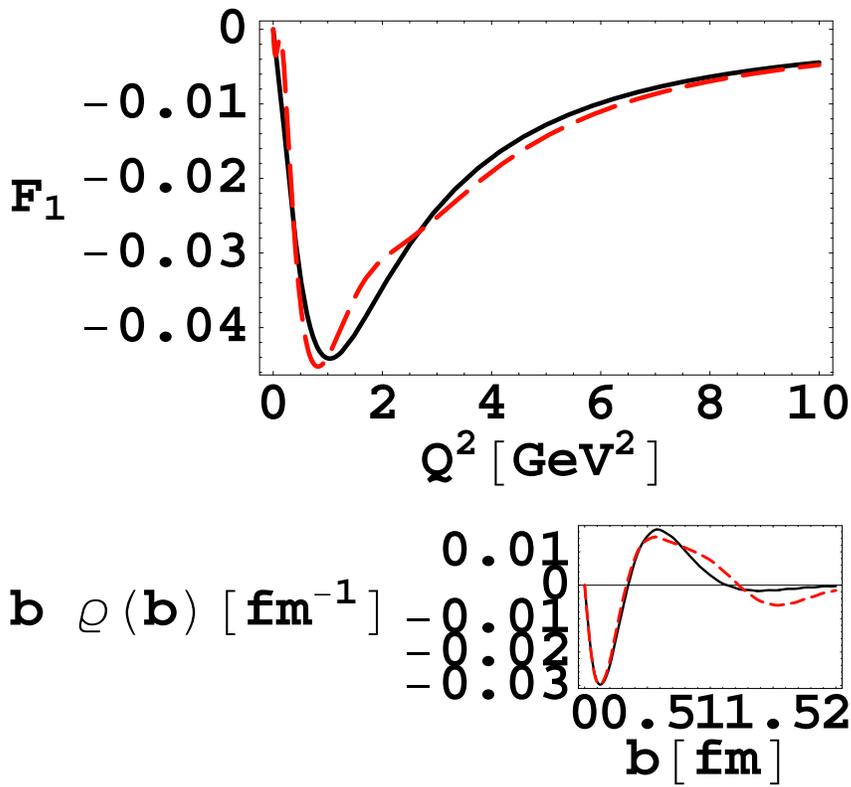}
\end{picture}
\caption{Upper panel: $F_1$. Lower panel:   $b \rho(b)$ in transverse position space. The
solid curves are  obtained using \cite{Kelly:2004hm} and the dashed curves with  \cite{Bradford:2006yz}.  }
\end{figure}
The negative central density deserves further explanation. 
The upper panel of Fig.~2 shows $F_1$ for the neutron obtained
using the  two different parameterizations which are observably different. However, in both
cases
 $F_1$ is  negative (because of the dominance of the $G_M$ term of \eq{use})
for all  values of
$Q^2$. This along with taking $b=0,\;J_0(Qb)=1$ in \eq{use} 
leads immediately to the central negative result. The long range
structure 
of the charge density is  captured by displaying the quantity $b\rho(b)$ in the lower panel of Fig.~2. 
At very large distances from the center,  again 
suggesting the existence of the  long-ranged pion cloud.

The present analysis provides 
 detailed information  about the location of charge density within the nucleon, and also
 incorporates the  lore regarding mean-square-radii, MSR. It has long
been  known that the  MSR defined by the form factor   $G_E$ is dominated
by the Foldy term $-1.91/(4M^2)= -0.126 \;{\rm fm}^{2}$ \cite{foldy, Meissner:2007tp}.
arising from the neutron magnetic moment $F_2(0)$. 
The experimental value of the $G_E$ MSR,  cited in  \cite{Kelly:2004hm}, is  (-0.114 $\pm$ 0.003) fm$^2$, so
 the MSR associated with $F_1$  (obtained from the
integral $\int d^2b b^2 \rho(b)$) is small and positive (+0.012 fm$^2$). This  result
is consistent with Figs.~1 and 2.  However, knowing the MRS of $F_1$ does not, by itself, allow
one to conclude that the central neutron charge density is negative, does not 
reveal the critical 
model-independent feature that at the very largest  distances the charge density is
negative and does not imply the oscillatory behavior displayed in Figs.~1 and 2.

One can gain information about the individual $u$ and $d$ quark densities by invoking charge symmetry (invariance under a  rotation by $\pi$ about the $z$ (charge) axis in isospin space) \cite{me} so that
the $u,d$ densities in the proton are the same as the $d,u$ densities in the neutron. We also neglect the effects of $s\bar{s}$ \cite{happex}
or heavier pairs
of quarks. In this case 
$\rho_u(b)=\rho_p(b)+\rho_n(b)/2,\;
\rho_d(b)=\rho_p(b)+2\rho_n(b).$ 
The results, shown in Fig.~3, and obtained with either form factor parameterization 
are that the central up quark density  is significantly larger than the central down
quark density. The latter becomes larger at distances away from the center.

\begin{figure}\label{neutronge}
\unitlength.8cm
\begin{picture}(14,8.2)(3.5,2.24)
\includegraphics{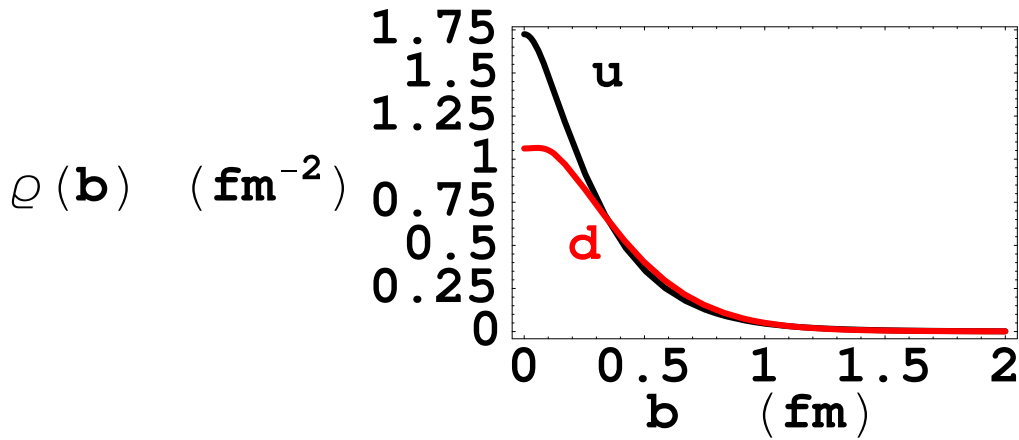}
\end{picture}
\caption{Transverse densities for up $u$ and down $d$ quarks. Each is normalized to unity} 
\end{figure}

Model independent information about parton distributions has been obtained. In particular, the central
density of the neutron is negative.
Future experimental measurements
of neutron electromagnetic form factors could render the present results more precise, or potentially modify
them considerably. Obtaining a quantitative and intuitive understanding of our results presents a challenge
to lattice QCD and to builders of  phenomenological models.

I thank the USDOE (FG02-97ER41014)
for partial support of this work, and 
 J. Arrington, A. Bernstein,  A. Bulgac, M. Burkardt, W. Detmold, S.~D.~Ellis, B. Holstein, and C. Hyde-Wright
for useful discussions.


\begin{thebibliography}{0}
\bibitem{reviews} 
 H.~y.~Gao,
  Int.\ J.\ Mod.\ Phys.\  E {\bf 12}, 1 (2003)
  [Erratum-ibid.\  E {\bf 12}, 567 (2003)];
 C.~E.~Hyde-Wright and K.~de Jager,
  Ann.\ Rev.\ Nucl.\ Part.\ Sci.\  {\bf 54}, 217 (2004);
  C.~F.~Perdrisat, V.~Punjabi and M.~Vanderhaeghen,
  arXiv:hep-ph/0612014.

\bibitem{soper1} D.E. Soper, Phys.\ Rev.\ D\ {\bf 15},
 1141 (1977). 

\bibitem{mbimpact}
  M.~Burkardt,
  Int.\ J.\ Mod.\ Phys.\  A {\bf 18}, 173 (2003).


\bibitem{diehl2} 
  M.~Diehl,
  Eur.\ Phys.\ J.\  C {\bf 25}, 223 (2002)
  [Erratum-ibid.\  C {\bf 31}, 277 (2003)].

\bibitem{notation}
Our notation is that $x^\pm\equiv (x^0\pm x^3)/\sqrt{2},p^\pm\equiv (p^0\pm p^3)/\sqrt{2}$, and
$p_\mu x^\mu=p^-x^++p^+x^--\bfp\cdot\bfb$.
The coordinates perpendicular to the 0 and 3 directions are denoted as
$\bfb$ and $\bfp$.

\bibitem{isgurk}
J.~L.~Friar,
  Part.\ Nucl.\  {\bf 4}, 153 (1972);
R.~D.~Carlitz, S.~D.~Ellis and R.~Savit,
  Phys.\ Lett.\  B {\bf 68}, 443 (1977);
  N.~Isgur, G.~Karl and D.~W.~L.~Sprung,
  Phys.\ Rev.\  D {\bf 23}, 163 (1981).
\bibitem{cbm}  A.~W.~Thomas, S.~Th\'eberge and G.~A.~Miller,
  Phys.\ Rev.\  D {\bf 24}, 216 (1981).
\bibitem{soper} J. Kogut and D.E. Soper, 
Phys.\ Rev.\ D\ {\bf 1}, 2901 (1970).


\bibitem{sachs} R. G. Sachs, Phys. Rev. {\bf 126}, 2256 (1962)
\bibitem{sachsexplain} 
The  momentum difference between  the
initial and final states 
appears via the use of  derivatives of momentum-conserving delta functions in the moments computed by
Ref.~\cite{sachs} 


\bibitem{vdh}  D. M\"uller et al., Fortschr. Phys. 
{\bf 42}, 101 (1994);
X.Ji, J.\ Phys.\ G\ {\bf 24}, 1181
(1998);
A.V. Radyushkin, Phys.\ Rev.\ D\ {\bf 56}, 5524
(1997);
K. Goeke, M.V. Polyakov, and 
M. Vanderhaeghen, Prog. Part. Nucl. Phys. {\bf 47}, 
401 (2001).

\bibitem{xdj} X. Ji, Phys.\ Rev.\ Lett. {\bf 78},
610 (1997).

\bibitem{scale} The  vector current $\bar{q}\gamma^\mu q$ is conserved,
so  $F_1$ is independent of the
 renormalization scale $\mu^2$
\cite{diehl2}. 

\bibitem{mb1} M. Burkardt, Phys.\ Rev.\ D\
{\bf 62}, 071503 (R) (2000).


\bibitem{diehl} M. Diehl {\it et al.}, Nucl.\ Phys.\
B\ {\bf 596}, {33} (2001).


\bibitem{Bradford:2006yz}
  R.~Bradford, A.~Bodek, H.~Budd and J.~Arrington,
  Nucl.\ Phys.\ Proc.\ Suppl.\  {\bf 159}, 127 (2006).

\bibitem{Kelly:2004hm}
  J.~J.~Kelly,
  Phys.\ Rev.\  C {\bf 70}, 068202 (2004).

\bibitem{Arrington:2003qk}
  J.~Arrington,
  Phys.\ Rev.\  C {\bf 69}, 022201 (R) (2004).
\bibitem {foldy} L.~L.~Foldy, Phys.\ Rev.\ {\bf83}, 688  (1951).
\bibitem{Meissner:2007tp}
  U.~G.~Meissner,
  AIP Conf.\ Proc.\  {\bf 904}, 142 (2007).
\bibitem{me}
 G.~A.~Miller, A.~K.~Opper and E.~J.~Stephenson,
  Ann.\ Rev.\ Nucl.\ Part.\ Sci.\  {\bf 56}, 253 (2006)
 G.~A.~Miller,
  Phys.\ Rev.\  C {\bf 57}, 1492 (1998);
G.~A.~Miller, A.~W.~Thomas and A.~G.~Williams,
  Phys.\ Rev.\ Lett.\  {\bf 56}, 2567 (1986).

\bibitem{happex} A.~Acha {\it et al.}, 
  Phys.\ Rev.\ Lett.\  {\bf 98}, 032301 (2007).


\end{thebibliography}
\end{document}